\documentclass[10pt, twocolumn]{IEEEtran}

\usepackage{multirow}
\usepackage{booktabs}
\usepackage{longtable}
\usepackage{cite}

\usepackage{graphicx}
\usepackage{amsmath}
\usepackage{amsfonts}
\usepackage{amssymb}
\usepackage{stfloats}
\usepackage{arydshln}

\usepackage{subfigure}
\graphicspath{{figure/}}

\newtheorem{remark}{Remark}

\usepackage{amsfonts, amsmath, mathrsfs, amsbsy, amssymb, dsfont, color}
\usepackage{graphicx}

\newcommand\prefixtext[1]{%
  \ifvmode\else\\\@empty\fi
  \noalign{%
    \penalty0%
    \vbox{\mathstrut}%
    \penalty10000%
    \vskip-\baselineskip
    \penalty10000%
    \vbox to 0pt{%
      \normalbaselines
      \ifdim\linewidth=\columnwidth
      \else
        \parshape\@ne
        \@totalleftmargin\linewidth
      \fi
      \vss
      \noindent#1\par}%
      \penalty10000%
      \vskip-\baselineskip}%
      \penalty10000}

\newtheorem{theorem}{Theorem}
\newtheorem{lemma}{Lemma}

\newcommand{\qed}{\nobreak \ifvmode \relax \else
      \ifdim\lastskip<1.5em \hskip-\lastskip
      \hskip1.5em plus0em minus0.5em \fi \nobreak
      \vrule height0.75em width0.5em depth0.25em\fi}

\DeclareMathAlphabet{\mathpzc}{OT1}{pzc}{m}{it}
\newcommand{\comment}[1]{}
\def\({\left(}
\def\){\left)}

\def\bth{\boldsymbol{\theta}}

\def\bnu{\boldsymbol{\nu}}

\def\bth{\boldsymbol{\theta}}

\def\bLa{\boldsymbol{\Lambda}}

\def\bSi{\boldsymbol{\Sigma}}

\def\tr{\text{tr}}
\def\({\left(}
\def\){\right)}
\def\[{\left[}
\def\]{\right]}
\def\BEq{\begin{eqnarray}}
\def\EEq{\end{eqnarray}}
\def\BE*{\begin{eqnarray*}}
\def\EE*{\end{eqnarray*}}
\def\BA{\begin{array}}
\def\EA{\end{array}}

\def\Nn{\nonumber}
\def\0{\mathbf{0}}
\def\1{\mathbf{1}}
\def\a{\mathbf{a}}
\def\A{\mathbf{A}}

\def\C{\mathbf{C}}

\def\F{\mathbf{F}}

\def\I{\mathbf{I}}

\def\l{\mathbf{l}}

\def\N{\mathbf{N}}

\def\P{\mathbf{P}}
\def\q{\mathbf{q}}

\def\S{\mathbf{S}}

\def\u{\mathbf{u}}

\def\w{\mathbf{w}}

\def\X{\mathbf{X}}
\def\y{\mathbf{y}}
\def\Y{\mathbf{Y}}
\def\z{\mathbf{z}}
\def\Z{\mathbf{Z}}

\def\and{\prefixtext{and}}

\def\diag{{\rm diag}}

\def\Im{\mathrm{Im}}
\def\Re{\mathrm{Re}}

\title{One-Bit Target Detection in Collocated MIMO Radar and Performance Degradation Analysis}
\author{Yu-Hang Xiao,~\IEEEmembership{Member,~IEEE}, David Ram{\'\i}rez,~\IEEEmembership{Senior Member,~IEEE}, Peter J. Schreier,~\IEEEmembership{Senior Member,~IEEE}, Cheng Qian,~\IEEEmembership{Member,~IEEE}, and Lei Huang,~\IEEEmembership{Senior Member,~IEEE}
\thanks{\textcolor{blue}{This work has been submitted to the IEEE for possible publication.
Copyright may be transferred without notice, after which this version may
no longer be accessible.}

Y.-H. Xiao is with the College of Electronics and Information Engineering, Shenzhen University,  Shenzhen 518060, China e-mail:
(yuhangxiao@szu.edu.cn).}
\thanks{David Ram{\'\i}rez is with the Department of Signal Theory and Communications, Universidad Carlos III de Madrid, Madrid 28903, Spain, and also with the Gregorio Mara{\~n}{\'o}n Health Research Institute, Madrid 28007, Spain (email: david.ramirez@uc3m.es).}
\thanks{P. J. Schreier is with the Signal and System Theory Group, Paderborn University, Paderborn 33098, Germany (e-mail:
 peter.schreier@sst.upb.de).}
\thanks{C. Qian is with the IQVIA Inc., Cambridge, MA 02139, USA (e-mail: alextoqc@gmail.com).}
\thanks{L. Huang is with the Guangdong Key Laboratory of Intelligent Information Processing and the Shenzhen Key Laboratory of Advanced Navigation Technology, Shenzhen University,  Shenzhen 518060, China (e-mail: dr.lei.huang@ieee.org).}
\thanks{The work of D. Ram{\'\i}rez was supported by the Ministerio de Ciencia, Innovaci{\'o}n y Universidades under grant TEC2017-92552-EXP (aMBITION), by the Ministerio de Ciencia, Innovaci{\'o}n y Universidades, jointly with the European Commission (ERDF), under grant TEC2017-86921-C2-2-R (CAIMAN), and by The Comunidad de Madrid under grant Y2018/TCS-4705 (PRACTICO-CM). The work of Lei Huang was supported in part by the National Science Fund for Distinguished Young Scholars under Grant 61925108, and in part by the National Natural Science Foundation
of China under Grants U1713217 and U1913221.}
}

\begin{document}

\input{epsf}
\date{}

\maketitle
\maketitle
\begin{abstract}


Target detection is an important problem in multiple-input multiple-output (MIMO) radar. Many existing target detection algorithms were proposed without taking into consideration the quantization error caused by analog-to-digital converters (ADCs).
This paper addresses the problem of target detection for MIMO radar with one-bit ADCs and derives a Rao's test-based detector. The proposed method has several appealing features: 1) it is a closed-form detector; 2) it allows us to handle sign measurements straightforwardly; 3) there are closed-form approximations of the detector's distributions, which allow us to theoretically evaluate its performance. Moreover, the closed-form distributions allow us to study the performance degradation due to the one-bit ADCs, yielding an approximate $2$ dB loss in the low-signal-to-noise-ratio (SNR) regime compared to $\infty$-bit ADCs. Simulation results are included to showcase the advantage of the proposed detector and validate the accuracy of the theoretical results.
\end{abstract}


\begin{sloppypar}
\section{Introduction}
Multiple-input multiple-output (MIMO) radar, which uses multiple antennas at both transmitter and receiver, can provide significant performance gains by exploiting waveform diversity. With the increase of array sizes and the emergence of resource-limited applications, one-bit sampling has become a promising technique, as it provides, on one hand, cost and energy efficiency, and, on the other, higher sampling rates.

During the past two decades, one-bit processing has been studied for many problems, ranging from direction-of-arrival (DOA) estimation~\cite{Shalom2002TAES,Yu2016SPL,Liu2017ICASSP}, MIMO communications~\cite{Qian2019SPL,Li2017TSP,Stein2018TSP,Choi2016TC}, frequency estimation~\cite{Madsen2000TSP} to target tracking~\cite{Stein2015TSP}. 
These works have shown that, with a rational design of signal processing algorithms, the performance degradation is usually relatively small. However, most of the existing algorithms in radar target detection were derived without considering the quantization effect~\cite{Liu2015TAES,Liu2018TSP,Xu2008TAES}. The performance degradation caused by imperfect analog-to-digital converters (ADCs) has also not been well studied in the literature.

The use of sign (or one-bit) measurements hinders the application of standard detection techniques, such as the generalized likelihood ratio test (GLRT), as in these cases, the likelihood function is a product of $Q$ functions, and it does therefore not admit a closed-form. Concretely, there does not exist analytical solution of the maximum likelihood estimates (MLE) of the unknown parameters, i.e., target reflectivity. Even though numerical optimization methods can be used to find the solution~\cite{Fang2013SPL}, they lead to detectors without closed form, which translates into a complicated performance analysis. An alternative to numerical methods was developed in~\cite{Cheng2019SPL}, which considered a simplified model. Concretely, it assumes that the target reflectivity is known \emph{a priori}. Nevertheless, this assumption is unrealistic from a practical standpoint since the reflectivity is fast-changing and generally needs to be estimated. In this paper, to avoid the computation of the MLE, we resort to Rao's test, which only requires the computation of the Fisher information matrix (FIM) and allows us to obtain a closed-form statistic. This also frees us from requiring prior information on the target reflectivity, as it has been implicitly estimated by a second-order Taylor's approximation of the MLE~\cite{Kay_detection}.

A very important problem is to analyze the performance degradation of the proposed Rao's test with one-bit ADCs compared to the $\infty-$bit case. It is worth mentioning that the performance loss analysis has been visited several times in the estimation literature. Despite the different backgrounds, it is quite commonly suggested that for symmetric one-bit ADCs, the performance loss is $\pi/2$ ($2$ dB) in the low-signal-to-noise-ratio (SNR) regime, while growing larger when the SNR increases~\cite{Stein2018TSP,Vleck1966,Stein2015TSP,Madsen2000TSP,Papadopoulos2001TIT}. However, this problem has attracted much less attention in the detection field. To the best of our knowledge, only~\cite{Fang2013SPL} proved a $2$ dB loss for all SNRs based on Wilks' theorem~\cite{Kay_detection}, which contradicts the general result in estimation papers. This is because Wilks' theorem is not generally applicable for one-bit quantized data, as will be shown later.

In this paper, we carry out the performance loss analysis by comparing the theoretical probabilities of false alarm and detection of the derived Rao's test with 1-bit ADCs and the GLRT with $\infty-$bit ADCs. Since, both, Rao's test and GLRT are asymptotically optimal~\cite{Kay_detection,Ghobadzadeh2016TSP}, their performances can be seen as the best-case scenario one can get with one-bit and $\infty-$bit ADCs. This motivates us to analyze the distribution of the detectors under the null and alternative hypotheses. We show that the proposed Rao's test is exponentially distributed in the null case. For the non-null distribution, we first provide a near-exact result based on Imhof's generalized non-central $\chi^2$ distribution~\cite{Imhof1961}, which is followed by a non-central $\chi^2$ approximation that is valid only in the low-SNR regime. For the $\infty-$bit case, the GLRT has an identical null distribution as its one-bit counterpart, whereas, in the non-null case, it is non-central $\chi^2$ distributed for all SNRs, with the non-centrality parameter increased by a scale of $\pi/2$. This proves that 
the $2$ dB performance degradation is achievable only in the low-SNR regime, which agrees better with the results in estimation papers.

The contributions of this paper are summarized as:
\begin{enumerate}
\item The Rao's test is formulated for one-bit detection in collocated MIMO radar. Compared to the GLRT, it has the advantage of being in closed-form, i.e., no iterative algorithms are required. In addition, it does not need prior information on the target reflectivity, as opposed to~\cite{Cheng2019SPL}.

\item We obtain near-exact null and non-null distributions of the devised detector. For the performance comparison, a low-SNR approximation of the non-null distribution is also provided, which shows that Wilks' theorem only works in the low-SNR regime, but not in general for one-bit quantized samples.

\item The performance of the detector is compared with the $\infty-$bit counterpart, which shows that the performance loss of using one-bit ADCs is as low as $2$ dB in the low-SNR regime. Alternatively, this performance gap could also be compensated by increasing the number of samples by a factor of $\pi/2$.

\end{enumerate}

The remainder of this paper is organized as follows. The signal model for one-bit detection in collocated MIMO radar is presented in Section \ref{sec:signal_model}. In Section \ref{sec:detector}, a detector based on Rao's test is derived, with its null and non-null distributions analyzed in Section \ref{sec:performance}. The performance degradation of using one-bit ADCs with respect to ideal ($\infty$-bit) ADCs is studied in Section \ref{sec:comparison}. Section \ref{sec:simulations} provides simulation results to validate the theoretical calculations. Finally, the main conclusions are summarized in Section \ref{sec:conclusions}.

\subsection*{Notation}

Throughout this paper, we use boldface uppercase letters for matrices, boldface lowercase letters for column vectors, and
light face lowercase letters for scalar quantities. The notation $\A\in\mathbb{R}^{p\times q} \ (\mathbb{C}^{p\times q})$ indicates that $\A$ is a $p\times q$ real (complex) matrix. The $(i,j)-$th entry of $\A$ is denoted by $\A_{ij}$, whereas $a_{i}$ corresponds to the $i-$th entry of the vector $\a$. The Frobenius norm and trace of $\A$ are $||\A||_F$ and $\tr(\A)$, and $\text{vec}(\A)$ is the vectorization of the matrix $\A$. The superscripts  $(\cdot)^{-1}$, $(\cdot)^T$ and $(\cdot)^H$ represent matrix inverse, transpose and Hermitian transpose operations. The operators $\mathbb{E}[a]$ and $\mathbb{V}[a]$ denote, respectively, the expected value and variance of $a$, $\mathbb{C}[a,b]$ is the covariance between $a$ and $b$, and $\sim$ means ``distributed as''. The $\chi^2_f$ and $\chi^2_f(\delta^2)$ denote, respectively, the central and non-central Chi-squared distributions, where $f$ is the number of degrees-of-freedom (DOF) and $\delta^2$ is the noncentrality parameter. Finally, the operators $\operatorname{Re}(\cdot)$ and $\operatorname{Im}(\cdot)$ extract the real and imaginary parts of their arguments, $\imath$ is the imaginary unit, and $\mathrm{sign}(\cdot)$ takes the sign of its argument.

\section{Signal model}
\label{sec:signal_model}

Consider a collocated MIMO radar system where there are $p$ transmit and $m$ receive antennas, which are collocated. The transmitter emits a probing beam towards the desired angle $\phi$.
Assuming the presence of a far field target, the received signal at the input of the ADCs can be written as
\begin{equation}
  \label{X}
\mathbf{X}=\beta \mathbf{a}_{r}(\phi) \mathbf{a}_{t}^{H}(\phi) \mathbf{S}+\mathbf{N},
\end{equation}
where $\mathbf{X} \in \mathbb{C}^{m \times n}$, with $n$ being the number of available snapshots, and $\N \in \mathbb{C}^{m \times n}$ is additive white Gaussian
noise \cite{Liu2021IET,Jin2020TAES,Liu2014TSP,Xi2020TAES,Xi2020TSP}. 
Here, $\beta$ is the unknown target reflectivity, i.e., a complex amplitude proportional to the radar cross section (RCS), $\mathbf{a}_{t}(\phi)\in\mathbb{C}^{p\times 1}$ and $\mathbf{a}_{r}(\phi)\in\mathbb{C}^{m\times 1}$ stand for the transmit and receive steering vectors\footnote{The steering vectors are known taking into account that $\phi$ and the array geometries are known.}, respectively, and $\S\in\mathbb{C}^{p\times n}$ is the known transmitted waveform with $\tr(\S\S^H)=n/p$.~
After one-bit quantization, the baseband signal becomes
\begin{align}\label{Y}
\Y=\mathcal{Q}(\X) = \mathrm{sign}(\Re(\X)) + \imath \mathrm{sign}(\Im(\X)),
\end{align}
where $\mathcal{Q}(\cdot)$ denotes the quantization operator.

Our task is to identify the presence or absence of the target based on the quantized observations $\Y$. Let $\mathcal{H}_1$ be the hypothesis where there is target in the received data, while $\mathcal{H}_0$ be the hypothesis where there is no target. Then, without quantization, the standard target detection problem can be described as the binary hypothesis test
\begin{align}\label{raw_model}
\begin{array}{l}
\mathcal{H}_{0}: \mathbf{X}=\mathbf{N}, \\
\mathcal{H}_{1}: \mathbf{X}=\beta \mathbf{a}_{r}(\phi) \mathbf{a}_{t}^{H}(\phi) \mathbf{S}+\mathbf{N},
\end{array}
\end{align}
whereas with one-bit ADCs, the target detection problem becomes
\begin{align}\label{quantized_model}
\begin{array}{l}
\mathcal{H}_{0}: \mathbf{Y}=\mathcal{Q}\left(\N\right), \\
\mathcal{H}_{1}: \mathbf{Y}=\mathcal{Q}\left(\beta \mathbf{a}_{r}(\phi) \mathbf{a}_{t}^{H}(\phi) \mathbf{S}+\mathbf{N} \right).
\end{array}
\end{align}

These two detection problems have been addressed in the past. The works in \cite{Liu2015TAES,Liu2018TSP,Xu2008TAES} have studied the target detection problem for MIMO radar without quantization in \eqref{raw_model}. Then,
\cite{Cheng2019SPL} considered the MIMO target detection problem from sign measurements, which can be viewed as a one-bit quantized version of that in \cite{Liu2015TAES,Liu2018TSP,Xu2008TAES}. However, the reflectivity parameter, $\beta$, was assumed to be known, which is usually unrealistic in practical MIMO radar systems. In this paper, we take $\beta$ as an unknown deterministic value, namely, no prior knowledge on $\beta$ is required. Mathematically, the above model can be seen as a complex version of the decentralized detection in wireless sensors network in~\cite{Fang2013SPL}, where the authors proposed a GLRT for real-valued measurements based on an iterative algorithm to seek the MLE of $\beta$. However, detectors without closed-form are less useful for performance analysis since only Wilks' theorem can be applied. In Section \ref{sec:performance} and \ref{sec:comparison}, we will show that Wilks' theorem only works for the low-SNR regime in one-bit detection. Therefore, to avoid the MLE computation, in this work we derive a closed-form detector based on Rao's test, which allows us to perform a more detailed performance assement.

\section{Derivation of Rao's Test}
\label{sec:detector}

The common approach to solve the above hypothesis testing problems with unknown parameters is to derive the GLRT. However, for one-bit measurements, the likelihood function is a product of $Q$ functions, which does not admit a closed-form expression, making the finding of the MLE of $\beta$ also nontrivial. This results in iterative detection algorithms, as in~\cite{Fang2013SPL}, and poses challenges for the subsequent performance analysis. Moreover, it can be seen in \eqref{quantized_model} that, after one-bit quantization, all amplitudes are lost, which means that  
the distribution of the sign measurements is independent of the noise power under $\mathcal{H}_0$, resulting a simple null hypothesis.

The aforementioned challenges motivate us to derive a detector based on Rao's test, as it does not require the MLE of the unknown parameters when $\mathcal{H}_0$ is simple. Before proceeding, and for notational simplicity, let us define $\Z = \mathbf{a}_{r}(\phi) \mathbf{a}_{t}^{H}(\phi) \mathbf{S}$, $\z = \text{vec}(\Z)$, $\y = \text{vec}(\Y)$ and
\begin{align}
z_i &= u_i + \imath v_i, & y_i &= r_i + \imath s_i.
\end{align}
with $i=1,\ldots,N$, where $N = mn$. Then, we have
\begin{align}
\beta z_i=a u_i- b v_i + \imath(av_i+bu_i),
\end{align}
where $\beta = a + \imath b$. Analogously to the real-valued case in~\cite{Fang2013SPL}, we can write the log-likelihood function under $\mathcal{H}_1$ as:
\begin{multline}\label{likelihood}
\mathcal{L}(\y;\bth)=\sum_{i=1}^{N}\log\(Q\[\frac{-r_i(au_i-bv_i)}{\sigma_n/\sqrt{2}}\]\)\\
+\sum_{i=1}^{N}\log\(Q\[\frac{-s_i(av_i+bu_i)}{\sigma_n/\sqrt{2}}\]\),
\end{multline}
where
\begin{align}
Q(x)=\int_x^{\infty}\frac{1}{\sqrt{2\pi}}\exp\(\frac{-x^2}{2}\)dx,
\end{align}
$\sigma_n^2$ is the noise variance and $\bth=[a,b,\sigma_n^2]^T$. Note that the log-likelihood under $\mathcal{H}_0$ could be easily obtained as $\mathcal{L}(\y;[\boldsymbol{0}^T, \sigma_n^2]^T)$. Due to the problem invariances, and as seen in \eqref{likelihood}, the likelihood of the data depends only on the ratio $\beta/\sigma_n^2$, instead of on $\beta$ and $\sigma_n^2$, individually.  Therefore, without loss of generality, we can simply set $\sigma_n^2=2$ and focus on $\beta$. That is, the real and imaginary parts of the noise both follow the standard Gaussian distribution, which yields $\bth_r=[a,b]^T$.

Now, the only unknown parameter is $\beta$ and it therefore follows that Rao's test statistic is given by~\cite{Kay_detection}
\begin{equation}\label{Rao_definition}
T_{\text{R}}=
\left(\!\left.\frac{\partial \mathcal{L}(\mathbf{y} ; \boldsymbol{\theta}_r)}{\partial \boldsymbol{\theta}_r}\right|_{\boldsymbol{\theta}_r=\boldsymbol{\theta}_{r,0}}
\!\right)^{T}\!\!\mathbf{F}^{-1}\!\!\left(\!\boldsymbol{\theta}_{r,0}\right)
\left(\left.\frac{\partial \mathcal{L}(\mathbf{y} ; \boldsymbol{\theta}_r)}{\partial \boldsymbol{\theta}_r}\right|_{\boldsymbol{\theta}_r=\boldsymbol{\theta}_{r,0}}
\!\right)
\end{equation}
where $\bth_{r,0}=[0,0]^T$ and $\F(\bth_r)$ is the Fisher information matrix (FIM):
\begin{equation}\label{FIM_definition}
\F(\bth_r)=\mathbb{E}\left[ \left.\left(\frac{\partial\mathcal{L}(\y;\bth_r)}{\partial \bth_r} \right)^{2} \right| \bth_r\right].
\end{equation}
The statistic in \eqref{Rao_definition} is obtained in the following theorem.

\begin{theorem}\label{theorem:RaoTest}
  The statistic of Rao's test is given by
  \begin{equation}
T_{\text{R}} =\frac{|\tr(\Y\Z^H)|^2}{N},
\end{equation}
  and the test is therefore
\begin{equation}\label{BiHyp}
  T_{\text{R}} \mathop{\gtrless}\limits_{\mathcal{H}_0}^{\mathcal{H}_1} \gamma,
\end{equation}
where $\gamma$ is a properly selected threshold.
\end{theorem}
\begin{IEEEproof}
  See Appendix \ref{appendix:A}.
\end{IEEEproof}

\section{Distributions of the proposed test}
\label{sec:performance}

In this section, we obtain approximate distributions of the proposed detector, $T_{\text{R}}$, under the null and alternative hypotheses. By computing the first and second order moments of the real and imaginary parts of the random variable $\tr(\Y\Z^H)$, a joint Gaussian approximation is established. Then, the distribution of $T_{\text{R}}$ is computed via a generalized non-central $\chi^2$ distribution. Since the computation of the first two moments under $\mathcal{H}_0$ can be obtained from those under $\mathcal{H}_1$, we will first address the non-null distribution.

\subsection{Distribution under $\mathcal{H}_1$}
Let us start by rewriting the test statistic as
\begin{align}
T_{\text{R}}=w_1^2+w_2^2
\end{align}
where
\begin{align}\label{w}
w_1&=\frac{\text{Re}(\z^H\y)}{\sqrt{N}}, &
w_2&=\frac{\text{Im}(\z^H\y)}{\sqrt{N}}.
\end{align}
We can now establish a bivariate Gaussian approximation for the joint distribution of $w_1$ and $w_2$. The result is stated
in the following theorem.
\begin{theorem}\label{theorem:1}
The distribution of $\w=[w_1,w_2]^T$ can be asymptotically ($n \rightarrow \infty$) approximated by the a bivariate real Gaussian distribution with mean
\begin{equation}\label{w_mean}
\u_w(\beta) = \frac{1}{\sqrt{N}}\begin{bmatrix}
\sum_{i=1}^{N}(c_i u_i + d_i v_i)\\ \\
\sum_{i=1}^{N}(d_i u_i - c_i v_i)
\end{bmatrix},
\end{equation}
and covariance matrix
\begin{equation}\label{w_covariance}
\bSi_w(\beta) =
\begin{bmatrix}
\sigma_{1}^2 & \sigma_{12}\\
\sigma_{21} & \sigma_{2}^2
\end{bmatrix},
\end{equation}
where
\begin{align}
\sigma_{1}^2 &=1-\frac{1}{N}\sum_{i=1}^{N}\left(c_i^2 u_i^2 + d_i^2 v_i^2\right),\\
\sigma_{2}^2 &=1-\frac{1}{N}\sum_{i=1}^{N}\left(c_i^2 v_i^2+ d_i^2 u_i^2\right),\\
\sigma_{12}&= \sigma_{21} = \frac{1}{N}\sum_{i=1}^{N}\left(c_i^2-d_i^2\right)u_iv_i, \label{sigma}
\end{align}
with
\begin{align}
c_i&=1-2Q(au_i-bv_i), &
d_i&=1-2Q(av_i+bu_i).
\end{align}
\end{theorem}

\begin{IEEEproof}
See Appendix \ref{appendix:B}.
\end{IEEEproof}

Theorem \ref{theorem:1} shows that the detector $T_{\text{R}}$ can be expressed as a squared sum of two correlated Gaussian random variables. To proceed, we need to use a linear transformation to convert them into two independent Gaussian random variables with different variances.
First, let us define\footnote{For notational simplicity, hereafter we drop the explicit dependency on $\beta$ of $\u_w(\beta)$ and $\bSi_w(\beta)$.}
\begin{equation}
\l=\bSi_w^{-\frac{1}{2}}(\w-\u_w).
\end{equation}
Then, we have
\begin{equation}
T_{\text{R}}=(\l+\bSi^{-\frac{1}{2}}\u_w)^T\bSi_w(\l+\bSi_w^{-\frac{1}{2}}\u_w).
\end{equation}
Using the eigenvalue decomposition of $\bSi_w$, given by $\bSi_w=\P^T\bLa\P$, where $\bLa=\diag(\lambda_1,\lambda_2)$, the test statistic can be rewritten as
\begin{align} \label{eq:Rao_weightedsum}
T_{\text{R}}
&=(\P\l+\P\bSi_w^{-\frac{1}{2}}\u_w)^T\bLa(\P\l+\P\bSi_w^{-\frac{1}{2}}\u_w)\Nn\\
&=\lambda_1 (\nu_1 + \mu_1)^2+\lambda_2 (\nu_2 + \mu_2)^2,
\end{align}
with
\begin{align}
\boldsymbol{\mu} &= \P\bSi_w^{-\frac{1}{2}}\u_w, & \bnu &=\P\l \sim \mathcal{N}(\boldsymbol{0},\I_2).
\end{align}
Thus, $T_{\text{R}}$ is distributed as the weighted sum of two indepedent non-central $\chi^2$ random variables with different centrality parameters, that is, a generalized non-central $\chi^2$ distribution. By integrating this probability density function (PDF) we could compute the probability of detection of the proposed Rao's test in Theorem \ref{theorem:RaoTest}, but there are no general expression for this PDF. Fortunately, the complementary distribution function of such a random variable, i.e., the required integral for computing the probability of detection, is given by Imhof~\cite{Imhof1961}, which is summarized in the following lemma.
\begin{lemma}
  \label{lem:Imhof}
Let $R=\sum_{r=1}^{l} k_{r} \chi^2_{h_{r}}(\delta_{r}^{2})$ be a weighted sum of non-central $\chi^2$ random variables with different centrality parameters and degrees of freedom. Then, its complementary distribution function is given by
\begin{align}\label{non_null}
\Pr\{R>x\}=\frac{1}{2}+\frac{1}{\pi} \int_{0}^{\infty} \frac{\sin \psi(u)}{u \rho(u)} d u,
\end{align}
where
\begin{equation}
\psi(u)=\frac{1}{2} \sum_{r=1}^{l}\left[h_{r} \tan ^{-1}\left(k_{r} u\right)+ \frac{\delta_{r}^{2} k_{r} u}{1+k_{r}^{2} u^{2}}\right]-\frac{1}{2} x u,
\end{equation}
and
\begin{equation}
\rho(u)=\prod_{r=1}^{l}\left(1+k_{r}^{2} u^{2}\right)^{\frac{1}{4} h_{r}} \times\exp \left\{\frac{1}{2} \sum_{r=1}^{l}\frac{\left(\delta_{r} k_{r} u\right)^{2}}{1+k_{r}^{2} u^{2}}\right\}.
\end{equation}
\end{lemma}

Then, we can use Lemma \ref{lem:Imhof} with $l=2$, $k_i=\lambda_i$, $\delta_i^2 = \mu_i^2,$ and $h_i=1$ to compute the probability of detection via numerical integration~\cite{Das2019}.

\subsection{Distribution under $\mathcal{H}_0$}

Under $\mathcal{H}_0$, we have
\begin{align}
\u_w&=\0, &
\bSi_w&=\I_2,
\end{align}
which implies that $w_1$ and $w_2$ are i.i.d. Gaussian random variables, yielding
\begin{equation}
T_\text{R}\sim \chi^2_2.
\end{equation}
Hence, $T_\text{R}$ is exponentially distributed with parameter $-1/2$, the probability of false alarm becomes
\begin{align}\label{eq:pfa}
P_{\rm fa}(\gamma) = \text{Pr}\{T_\text{R}>\gamma\}=\exp(-\gamma/2),
\end{align}
and the detection threshold can be obtained as follows
\begin{align}
\gamma=-2\log(P_{\rm fa}).
\end{align}

\section{Analysis of the Performance Degradation}
\label{sec:comparison}
In this section, we study the detection performance degradation of using one-bit ADCs with respect to
$\infty$-bit ADCs.
Since both the Rao's test and the GLRT are asymptotically optimal detectors \cite{Ghobadzadeh2016TSP}, their performances can be taken as the best achievable results in either one-bit or $\infty$-bit scenarios. Therefore, we can compute the performance degradation by comparing the performance of the Rao's test for one-bit case and the GLRT for $\infty$-bit case.
This aforementioned analysis could be achieved by using a series of analytical tools such as moment-based method~\cite{Wei2012a} or the asymptotic expansion method~\cite{test_of_independence,test_of_sphericity}, which could analyze the performance of detectors in a very accurate manner. However, to make a trade-off between the simplicity and accuracy of the expressions, here, we choose Wilks' theorem, which provides a simple expression for the performance metrics and allows us to gain insights.

\subsection{$\infty$-bit Case}

Let us first study the performance of the GLRT, $T_\mathrm{GLRT}$, with $\infty$-bit ADCs. In this case, there is no quantization error, so the received signal is $\X$. Then, we must solve the detection problem in \eqref{raw_model} whose GLRT can be obtained in a similar manner to \cite{Xu2008TAES} and is presented next..
\begin{theorem}
  \label{theorem:GLRT}
  The GLRT for the test in \eqref{raw_model} is
\begin{equation}\label{GLRT}
T_{\text{GLRT}} =\[1-\frac{|\tr(\X\Z^H)|^2}{N \tr(\X\X^H)}\]^{N} \mathop{\gtrless}\limits_{\mathcal{H}_1}^{\mathcal{H}_0} \gamma,
\end{equation}
where $\gamma$ is a properly selected threshold.
\end{theorem}
\begin{IEEEproof}
See~\cite{Xu2008TAES}. 
\end{IEEEproof}

It is easy to show that \eqref{GLRT} is a monotone transformation of
\begin{equation}
T_{\text{GLRT}}'=\frac{|\tr(\X\Z^H)|^2}{\tr(\X\X^H)},
\end{equation}
which is equivalent to $T_{\text{R}}$ when $\X$ is replaced by the one-bit quantized data $\Y$ and also noting that $\tr(\Y\Y^H)=2N$. This shows that the same detector can be applied to both one-bit and $\infty-$bit quantized data.

Once we have the GLRT, we can use Wilks' theorem~\cite[6C.1]{Kay_detection}, which is presented next, to obtain its asymptotic distributions.
\begin{lemma}\label{lemma:Wilks}
Consider the binary hypothesis testing problem
\begin{align*}
&\mathcal{H}_0: \bth = \bth_{r_0},\bth_{s},\\
&\mathcal{H}_1: \bth \neq \bth_{r_0},\bth_{s},
\end{align*}
where $\bth_{s}\in\mathbb{R}^{g\times1}$ is the nuisance parameter vector and $\bth_{r_0}\in\mathbb{R}^{f\times1}$. Then, the GLRT is asymptotically distributed as
\begin{equation}
  \label{eq:distributions}
  -2\log(T_{\textrm{\emph{GLRT}}}) \sim \begin{cases}
\chi^2_f, & \text{under}~\mathcal{H}_0,\\
\chi^2_f(\delta^2), &\text{under}~\mathcal{H}_1, \\
\end{cases}
\end{equation}
where
\begin{align}
    \delta^2 =(\bth_{r_1}-\bth_{r_0})^T\[\F^{-1}(\bth_{1})\]_{\bth_{r},\bth_{r}}^{-1}(\bth_{r_1}-\bth_{r_0}),
\end{align}
with $\bth_1=(\bth_{r_1}^T,\bth_s^T)^T$ being the true parameters in $\mathcal{H}_1$.
\end{lemma}

It is easy to see that the nuisance parameter is the noise variance, i.e., $\bth_s=\sigma_n^2$. Hence, since $\bth=[a,b,\sigma_n^2]^T$, it is obvious that the DOF is $f=2$. Moreover, the non-centrality parameter of the non-null distribution is
\begin{equation}\label{non_centrality_parameter}
\delta^2 = N|\beta|^2,
\end{equation}
which is derived in Appendix \ref{appendix:D}.

\subsection{One-bit case}

Since Rao's test has the asymptotic performance of the GLRT, intuitively, one could also apply Wilks' theorem for $T_\text{R}$, which would result in a similar non-central $\chi^2$ approximation. Thus, the performance degradation could be easily attained by a comparison between the non-centrality parameters~\cite{Fang2013SPL}. However, unlike the $\infty-$bit case, Wilks' theorem does not generally hold for one-bit quantized data due to the strong non-linearity of the ADCs. More specifically, as shown in \eqref{sigma}, $w_1$ and $w_2$ in $T_{\text{R}}$ are correlated, which makes the non-central $\chi^2$ approximation invalid, requiring the more sophisticated generalized non-central $\chi^2$ distribution. Nevertheless, we can still provide a non-central $\chi^2$ approximation to the non-null distribution in the low-SNR regime. The result is summarized in the following theorem.
\begin{theorem}
\label{th:nonnull_approx_Rao}
When the magnitude of $\beta$ is of order $\mathcal{O}(N^{-1/2})$, the distribution under $\mathcal{H}_1$ of $T_{\text{R}}$ can be approximated as
\begin{equation}\label{non_null_low_snr}
T_{\text{R}}\sim \chi^2_2(\delta^2_1),
\end{equation}
where
\begin{equation}\label{sigma_low_snr}
\delta^2_1=\frac{2N}{\pi}|\beta|^2.
\end{equation}
\end{theorem}
\begin{IEEEproof}
See Appendix \ref{appendix:E}.
\end{IEEEproof}

Comparing \eqref{sigma_low_snr} with \eqref{non_centrality_parameter}, and taking into account that the null distributions of $T_{\text{GLRT}}$ and $T_{\text{R}}$ are identical, the performance degradation in the low-SNR regime is about $ 10 \log_{10}(\pi/2)\approx 2$ dB.

\begin{remark}
In many parameter estimation problems, e.g., \cite{Stein2018TSP,Vleck1966,Stein2015TSP,Madsen2000TSP,Papadopoulos2001TIT}, the authors have studied the performance degradation by comparing the best achievable performance of an estimator with the Cram\'er-Rao bound, showing that the minimum achievable loss was also $2$ dB in the low-SNR regime. This matches our result in the detection problem. Nevertheless, one remarkable difference is that in the estimation case, the Fisher information is proportional to the performance bound  \textcolor{black}{and differ by the} ratio of $2/\pi$ in the low-SNR regime only. In our problem, the FIMs have a fixed gap of $2/\pi$, but the detection performances are not, in general, proportional to the Fisher information matrix. The underlying reason for this phenomenon will be an interesting future work.
\end{remark}

\section{Numerical Results}
\label{sec:simulations}

In this section, we carry out numerical simulations to validate our theoretical findings. We first evaluate the accuracy of the derived theoretical probabilities of false alarm and detection, which includes one-bit and $\infty-$bit scenarios, in the low-SNR and high-SNR cases. Then, we illustrate the detection performance and verify the $2$ dB performance degradation. All results are obtained from $10^6$ Monte Carlo trials.

We consider a collocated MIMO radar system with uniform linear arrays with half-wavelength inter-element spacing. Similar to \cite{Cui2014TSP,Aldayel2016}, we choose the orthogonal linear frequency modulation (LFM) signal as the transmitted waveform, that is,
\begin{align}
\mathbf{S}_{k, l}=\frac{\exp \left\{\imath 2 \pi k(l-1) / n+\imath \pi(l-1)^{2} / n\right\}}{\sqrt{p}},
\end{align}
where $k=1,\ldots,p$ and $l=1,\ldots,n$. Unless otherwise stated, the DOA $\theta$ is fixed at $-\pi/3$. The noise is defined as white Gaussian noise with zero mean and variance $\sigma_n^2 = 2$, 
and $\beta$ is generated from a complex Gaussian distribution with zero mean and unit variance, which is scaled to achieve the desired SNR, defined as:
\begin{equation}
\textrm{SNR}=10\log_{10}\!\(\frac{|\beta|^2}{\sigma_n^2}\).
\end{equation}
To quantify the approximation error between the theoretical and empirical cumulative distribution functions (CDFs), we shall also use the Cram\'{e}r-von Mises goodness-of-fit test, which is defined as~\cite{Anderson1962a}
\begin{equation}\label{Error}
  \epsilon = \frac{1}{K}\sum_{i=1}^K\left|F(c_i) - \hat{F}(c_i)\right|^2,
\end{equation}
where $F(c_i)$ is the empirical CDF and $\hat{F}(c_i)$ is the proposed approximation.

\subsection{Null Distribution}

\begin{figure}[!htbp]
\begin{minipage}[b]{1\linewidth}
  \centering{\includegraphics[width=0.9\columnwidth]{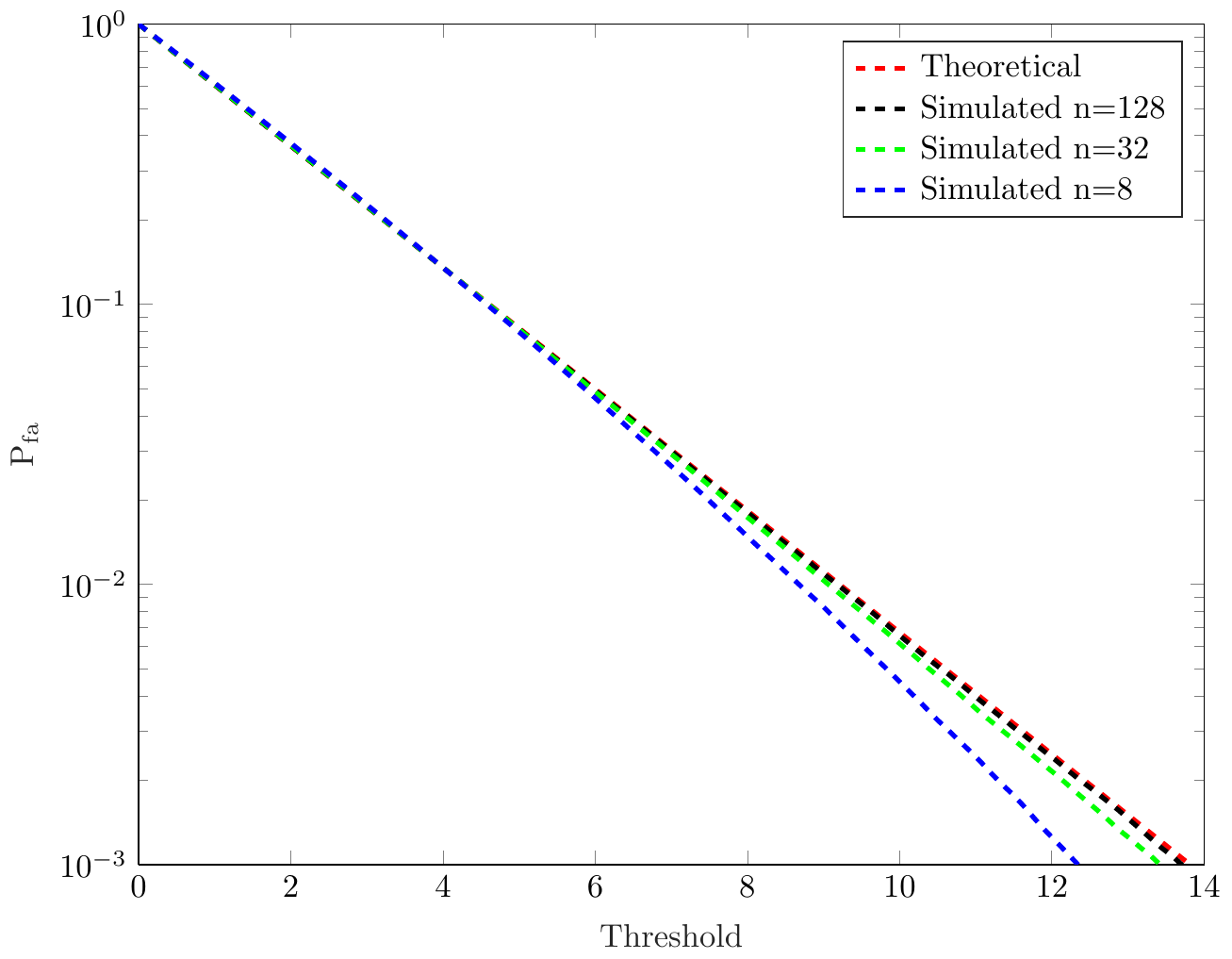}}
\centerline{\small{(a) Rao's test (one-bit data)}}
\medskip
\end{minipage}
\hfill
\begin{minipage}[b]{1\linewidth}
  \centering{\includegraphics[width=0.9\columnwidth]{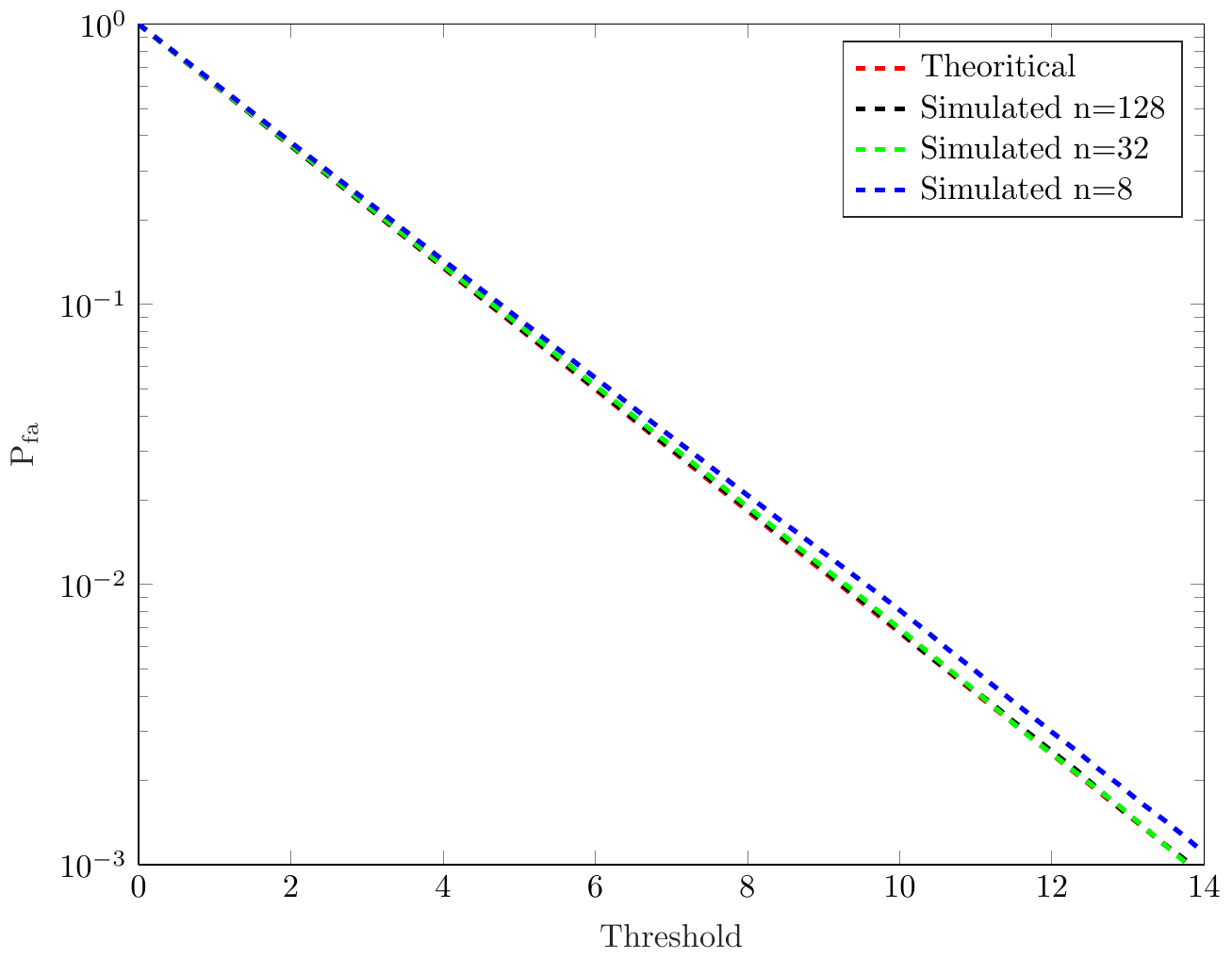}}
\centerline{\small{(b) GLRT ($\infty$-bit data)}}
\medskip
\end{minipage}
\caption{Probability of false alarm versus threshold for $m=p=4$ and $n=8,32,$ and $128$.}\label{fig1}
\end{figure}

First, we study the null distribution of the proposed Rao's test for one-bit ADCs and the GLRT for $\infty$-bit ADCs. Concretely, we used these distributions to compute the probabilities of false alarm, $P_{\rm fa}$. Fig. \ref{fig1}(a) depicts $P_{\rm fa}$ for $m=p=4$ as a function of the threshold for the approximation in \eqref{eq:pfa}, which is valid for large $N$ or, equivalently, $n$. This figure also shows the empirical probabilities obtained using Monte Carlo simulations with $n = 8, 32,$ and $128$, which allows us to see the accuracy of the proposed approximation for values of $n$ as small as $32$. Additionally, we have also computed the approximation errors, $\epsilon$, between the approximated and empirical CDFs, which are given by \textcolor{black}{$\epsilon = 1.75 \times 10^{-5}$ for $n = 8$, $\epsilon = 0.99 \times 10^{-7}$ for $n = 32$, and $\epsilon = 7.38 \times 10^{-8}$ for $n = 128$.} This confirms the asymptotic nature of the proposed approximation, which is due to the asymptotic Gaussian approximation.

In Fig. \ref{fig1}(b), for the same scenario as before, we exam the accuracy of the null distribution of $T_{\text{GLRT}}$. The approximated and empirical distributions also agree very well with each other, with approximation errors as small as \textcolor{black}{$\epsilon = 3.40 \times 10^{-5}$ for $n = 8$, $\epsilon = 2.52 \times 10^{-6}$ for $n = 32$, and $\epsilon = 3.06 \times 10^{-7}$ for $n = 128$.}

\subsection{Non-null Distribution}

\begin{figure}[!htbp]
\begin{minipage}[b]{1\linewidth}
  \centering{\includegraphics[width=0.9\columnwidth]{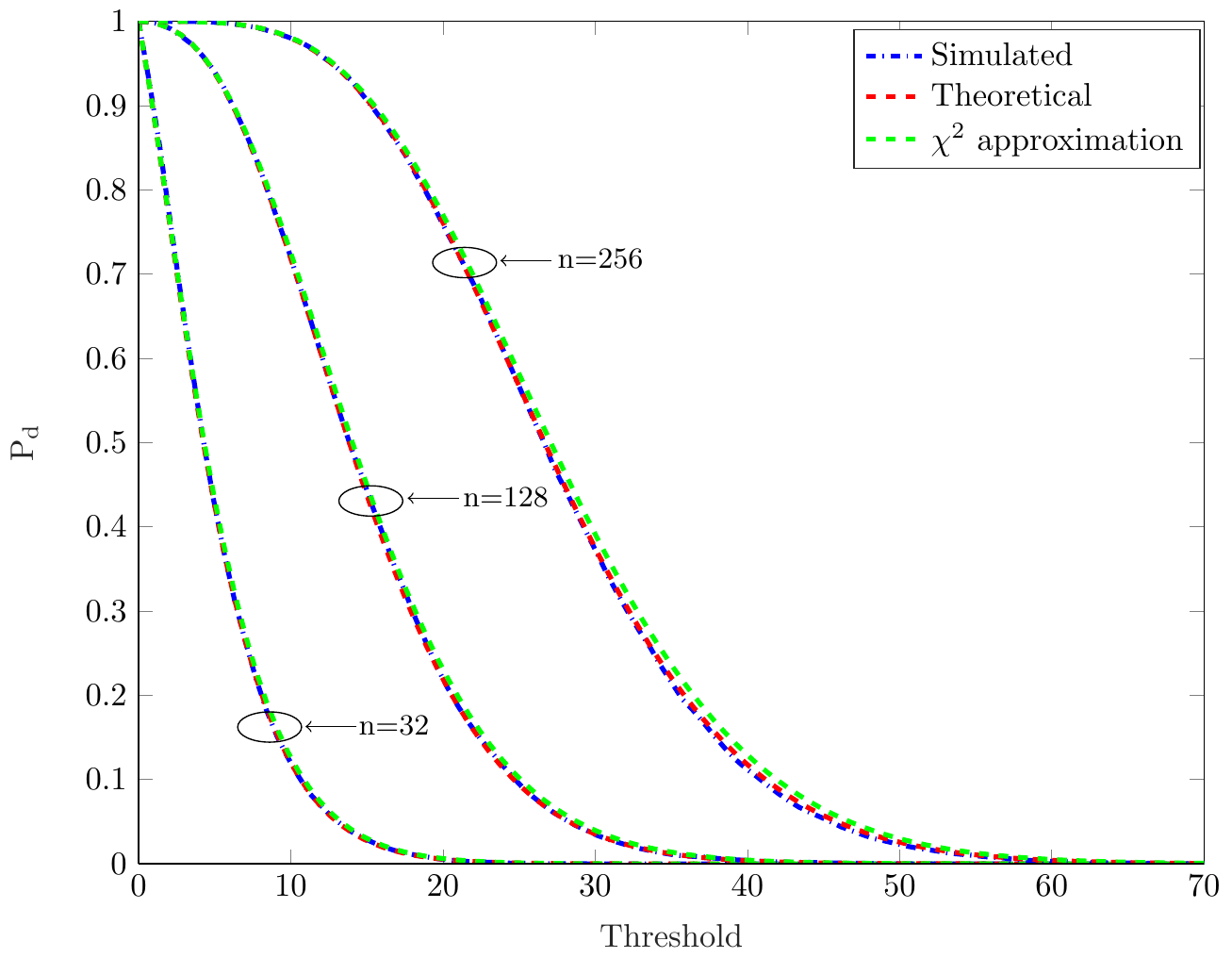}}
\centerline{\small{(a) \textcolor{black}{$\text{SNR} =-17$dB}}}
\medskip
\end{minipage}
\hfill
\begin{minipage}[b]{1\linewidth}
  \centering{\includegraphics[width=0.9\columnwidth]{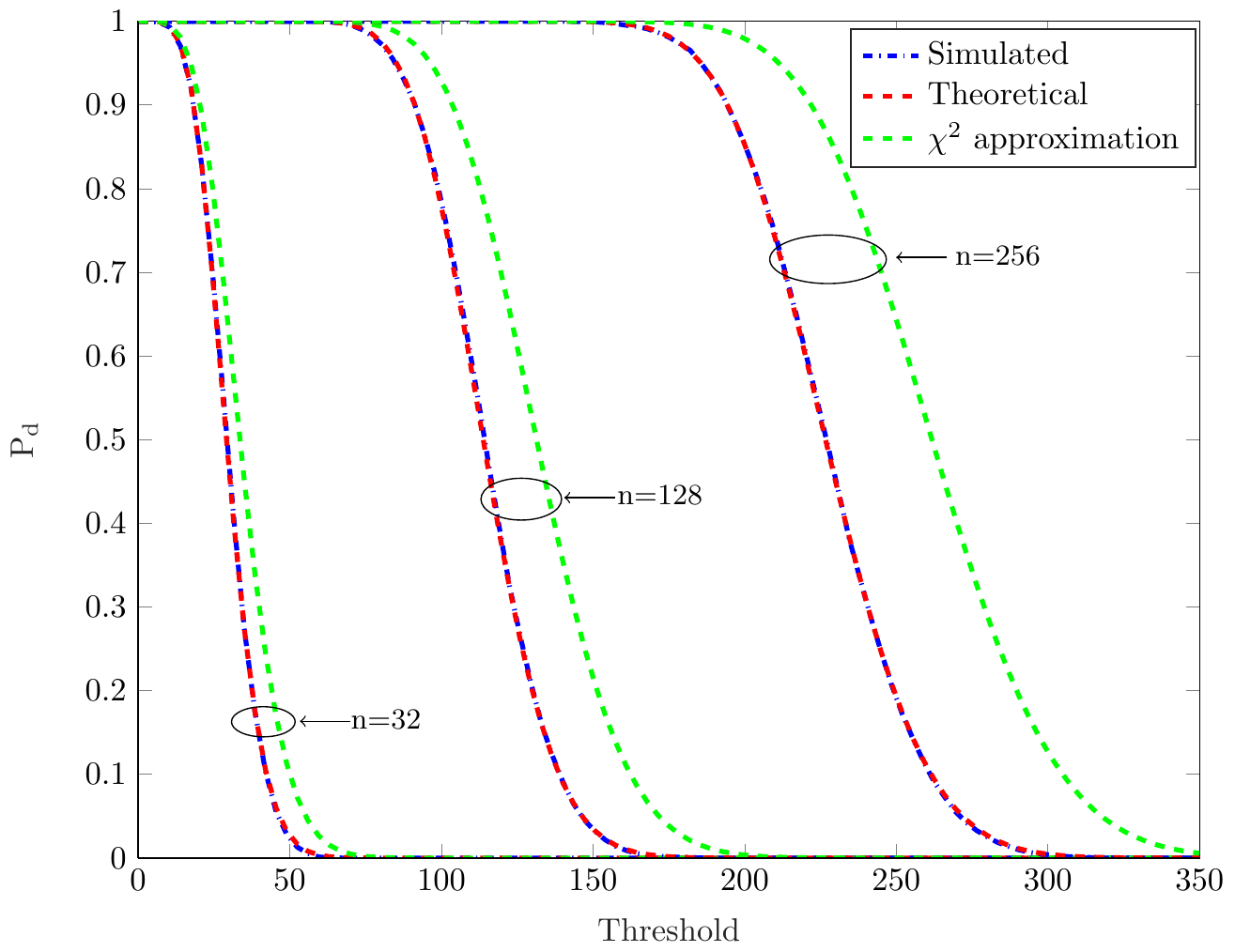}}
\centerline{\small{(b) \textcolor{black}{$\text{SNR} =-7$dB}}}
\medskip
\end{minipage}
\caption{Probability of detection of Rao's test (one-bit data) versus threshold for $m=p=4$ and $n=32,128,$ and $256$.}\label{fig2}
\end{figure}

This section studies the accuracy of the approximations of the probability of detection, including \eqref{non_null} and \eqref{non_null_low_snr} for Rao's test and \eqref{eq:distributions} for the GLRT. We first examine the accuracy of the distributions for one-bit data, namely \eqref{non_null} for the general case and \eqref{non_null_low_snr} for the low-SNR regime. The results are shown in Fig. \ref{fig2}, where the experiment parameters are $m=p=4$, $n=32,128,$ and $256$, and $\text{SNR}=-23$ dB for Fig. \ref{fig2}(a) and $\text{SNR}=-13$ dB for Fig. \ref{fig2}(b). The approximation errors are also summarized in Table \ref{tab:errors}, which shows that the analytic result in \eqref{non_null} works very well for both cases, while \eqref{non_null_low_snr} is accurate in the low-SNR regime.

\begin{table*}[!htbp]
\begin{center}
\caption{Errors of Different Approximations Methods at Different SNRs.}\label{tab:errors}
\begin{tabular}{ccc}
   \toprule[1.5pt]
   & $\text{SNR}=-23$ dB, $m=p=4$ & $\text{SNR}=-13$ dB, $m=p=4$ \\\midrule[1pt]
  Approximation &~\!\! $n=32$ ~~~~~~~~~~~\!$n=128$ ~~~~~~~~~~\!\! $n=256$~~~ & $n=32$ ~~~~~~~~~~ $n=128$ ~~~~~~~~ $n=256$ \\\midrule[1pt]
   \!\!Eq.~\eqref{non_null} & ~~~$1.89\times10^{-6}$ ~~~ $1.89\times10^{-7}$ ~~~~$3.03\times10^{-7}$~~~~\!&~~$3.91\times10^{-6}$    ~~~~ $1.53\times10^{-6}$ ~~~~$1.15\times10^{-6}$ \\
   \!\!Eq.~\eqref{non_null_low_snr} & ~~~$1.89\times10^{-5}$ ~~~ $9.19\times10^{-5}$ ~~~~$2.23\times10^{-4}$~~~~\!&~~$4.24\times10^{-3}$     ~~~~ $2.86\times10^{-2}$ ~~~~$2.86\times10^{-2}$ \\  \bottomrule[1.5pt]
\end{tabular}
\end{center}
\end{table*}

\begin{figure}[htbp]
  \centering{\includegraphics[width=0.9\columnwidth]{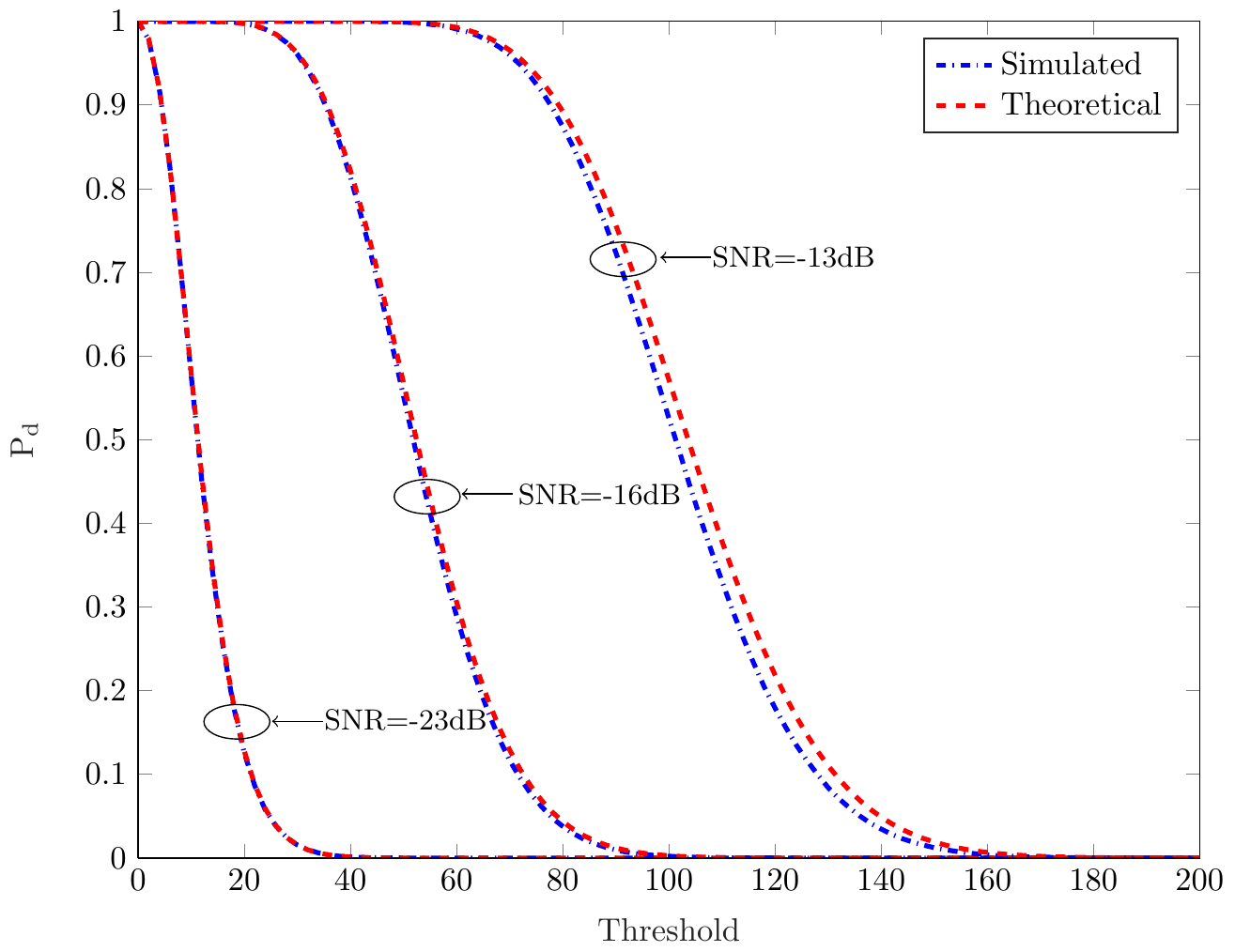}}
\caption{Probability of detection of the GLRT ($\infty$-bit data) versus threshold for $m=p=4$,  \textcolor{black}{$n = 256$} and $\text{SNR} = -13, -16,$ and $-23$ dB.}\label{fig3}
\end{figure}

In Fig. \ref{fig3}, we check the accuracy of the non-null distribution in \eqref{eq:distributions}, which corresponds to the GLRT with $\infty$-bit ADCs. Here, we investigate both high and low SNR regimes with $m=p=4$ and $n = 64$. Concretely, we have considered \textcolor{black}{$\rm SNR=-13\rm dB$, $\rm SNR=-16\rm dB$, and $\rm SNR=-23\rm dB$, which yield the errors $\epsilon = 8.53 \times 10^{-7}$, $\epsilon = 3.89 \times 10^{-5}$, and $\epsilon = 4.16 \times 10^{-4}$.} These results confirm that the derived non-null distributions are able to accurately predict the detection performance and enables us to compare the performance degradation of one-bit and $\infty$-bit ADCs theoretically, as done in Section \ref{sec:comparison}.

\subsection{Performance Degradation}

In this section, we study the probability of detection, for a fixed probability of false alarm, of the proposed detector $T_{\text{R}}$ and verify the prediction of $2$ dB performance degradation. It is worth mentioning that~\cite{Cheng2019SPL} has also considered the one-bit MIMO radar detection problem, but adopted a simplified model whereby the reflection parameter $\beta$ is {\it known}.
The likelihood ratio test (LRT) in~\cite{Cheng2019SPL} is:
\begin{multline}
T_{\text{LR}}=\sum_{i=1}^{N}\log\(Q\[{-r_i(au_i-bv_i)}\]\)\\
+\sum_{i=1}^{N}\log\(Q\[{-s_i(av_i+bu_i)}\]\)+2N\log(2),
\end{multline}
However, it should be noted that in real-world applications $\beta$ can hardly be known. Therefore, the above LRT is only used as a benchmark, but cannot be applied in realistic conditions.

\begin{figure}[!htbp]
  \centering{\includegraphics[width=0.9\columnwidth]{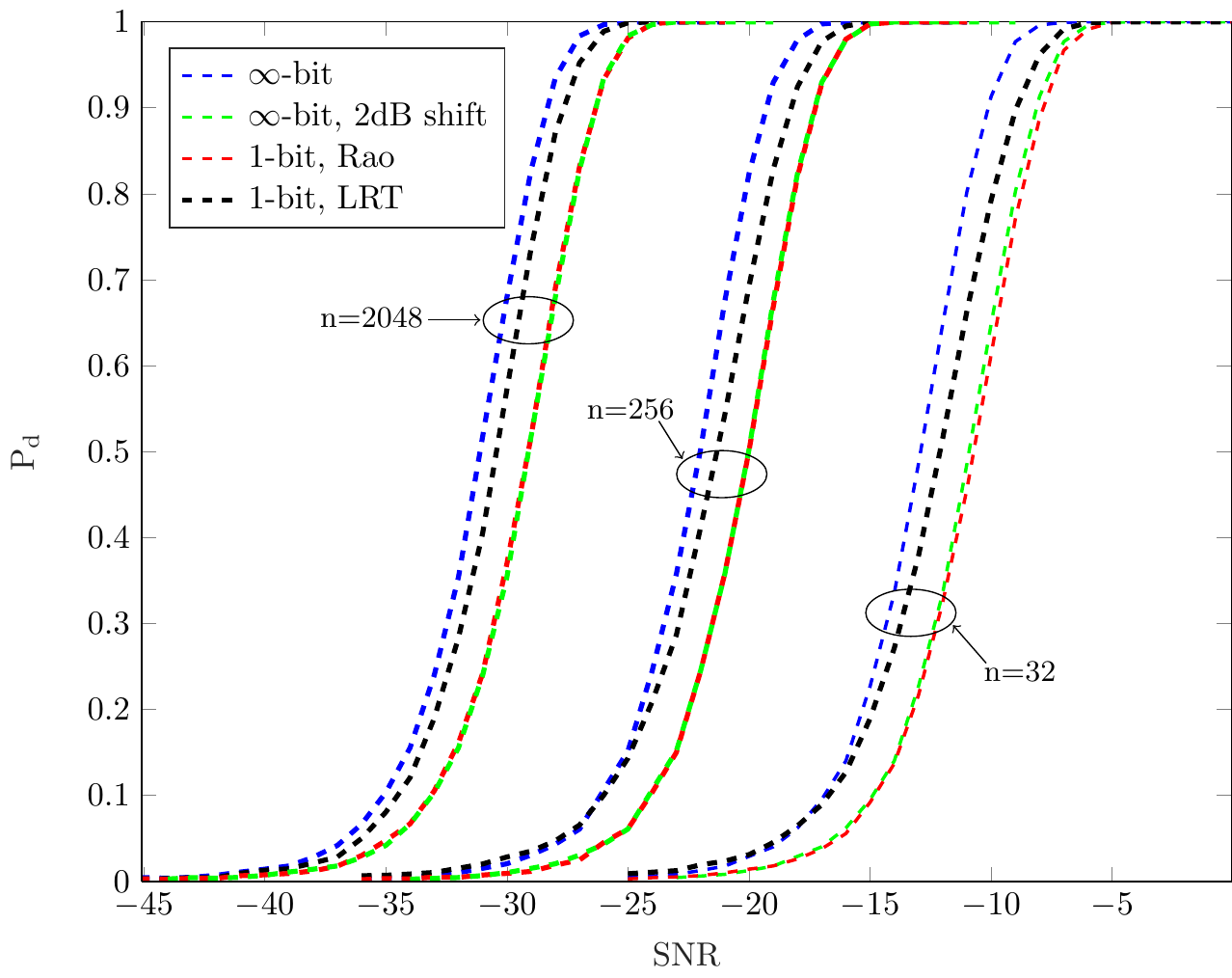}}
\caption{Probability of detection versus the SNR for $P_{\rm fa} = 10^{-3}$, $m=p=4$ and $n = 32, 256,$ and $2048$.} \label{fig:pd_SNR}
\end{figure}

Fig. \ref{fig:pd_SNR} depicts the probability of detection versus the SNR for $P_{\rm fa} = 10^{-3}$, $m=p=4$ and $n = 32, 256,$ and $2048$. This figure shows that the performance of the detector in the one-bit case is $2$ dB away from the case of $\infty-$bit quantized data, which matches very well with the theoretical prediction. Note that this approximation also holds for moderate SNRs, despite it was derived for low SNRs. In addition, the distances between $T_{\text{R}}$ and $T_{\text{LR}}$ are roughly $2$ dB when SNR is low but narrows to $1$ dB as $P_d$ approaches $1$. This is because for large SNRs, the estimate of $\beta$ is more accurate and the prior information has therefore less impact on the detection performance.

\begin{figure}[!htbp]
  \centering{\includegraphics[width=0.9\columnwidth]{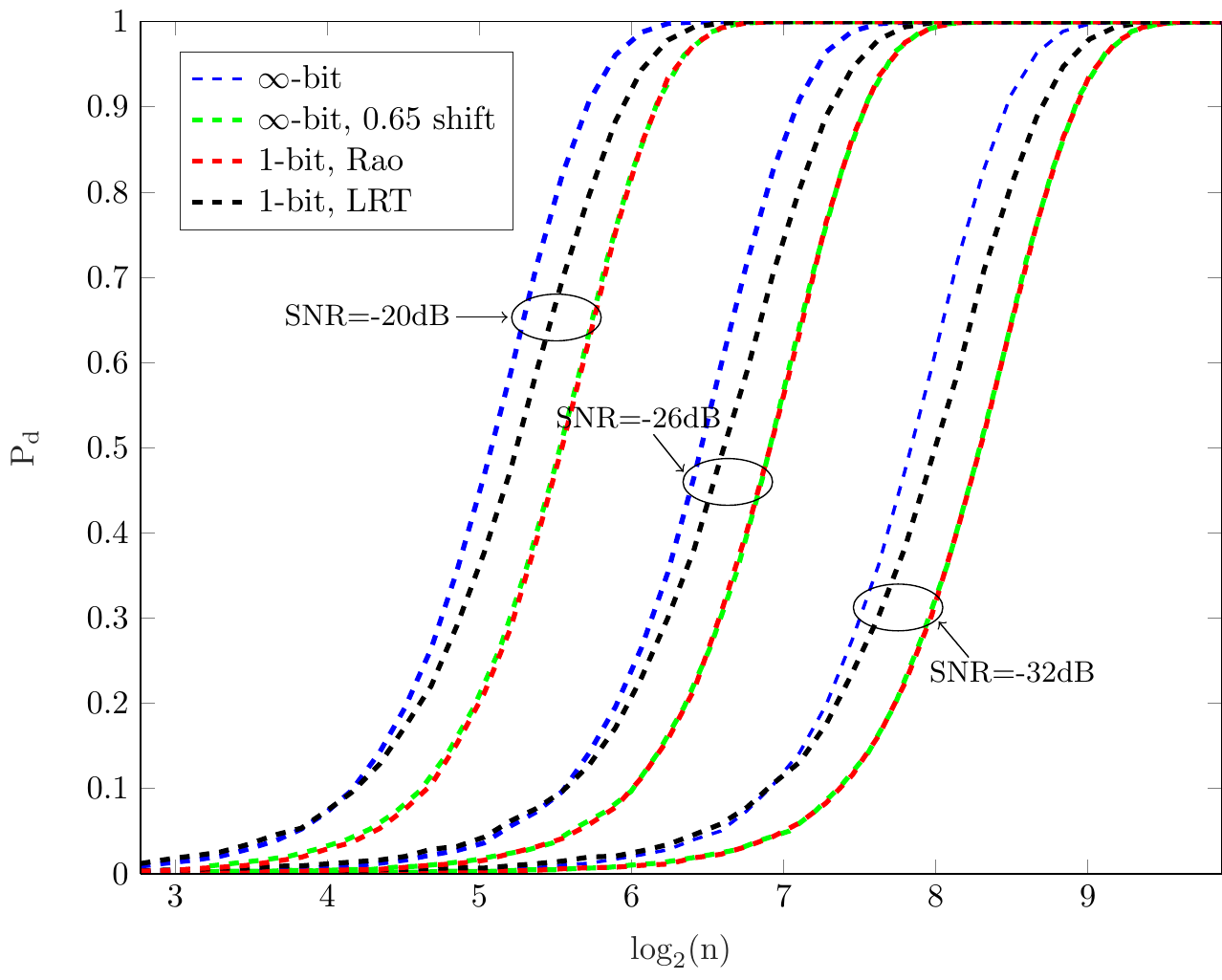}}
\caption{Probability of detection versus $n$ for $P_{\rm fa} = 10^{-3}$, $m=p=4$ and $\text{SNR} = -26, -32,$ and $-38$ dB.}
\label{fig:pd_n}
\end{figure}

Comparing \eqref{sigma_low_snr} with \eqref{non_centrality_parameter}, it is easy to see that the 2dB performance degradation can also be compensated by a $\pi/2$ multiplication of the number of samples. In Fig. \ref{fig:pd_n}, we plot the probability of detection versus number of samples $n$, in logarithmic scale, for an experiment with $m=p=4$ and several SNR values. To better illustrate the performance gap, a copy of the curve of the GLRT with $\infty$-bit ADCs is shifted by $\log_2(\pi/2) \approx 0.65$. It can be observed that the curve of the proposed Rao's test matches perfectly the shifted one, proving that performance loss can also be compensated by $\pi/2$ times amount of samples.

\section{Conclusion}
\label{sec:conclusions}

In this paper, we derived a novel detector based on Rao's test for target detection in MIMO radar with one-bit ADCs. The proposed method has a closed-form expression and does not require complicated numerical optimization of the reflectivity. We provided a comprehensive analysis of the theoretical performance of the proposed detector by deriving closed-form approximations for the null and non-null distributions, which allow us to compute the probabilities of false alarm and detection. Furthermore, we studied the performance degradation of using one-bit ADCs by comparing it with the performance of the GLRT with $\infty$-bit ADCs. Simulation results validated the effectiveness of the proposed detector and the related theoretical analysis.

\appendices

\section{Proof of Theorem \ref{theorem:RaoTest}}
\label{appendix:A}

We shall start by computing the derivative of the likelihood with respect to the unknown parameters, which are given by
\begin{multline}\label{derivative_1}
\frac{\partial\mathcal{L}(\y;\bth_r)}{\partial a}=\sum_{i=1}^{N}\frac{r_i u_i \varphi[r_i(au_i-bv_i)]}{Q[-r_i(au_i-bv_i)]}
\\ +\sum_{i=1}^{N}\frac{s_i v_i\varphi[s_i(av_i+bu_i)]}{Q[-s_i(av_i+bu_i)]},
\end{multline}
and
\begin{multline}\label{derivative_2}
\frac{\partial\mathcal{L}(\y;\bth_r)}{\partial b}=-\sum_{i=1}^{N}\frac{r_i v_i\varphi[r_i(au_i-bv_i)]}{Q[-r_i(au_i-bv_i)]}
\\ +\sum_{i=1}^{N}\frac{s_i u_i\varphi[s_i(av_i+bu_i)]}{Q[s_i(av_i+bu_i)]},
\end{multline}
where $\varphi(\cdot)$ is the standard Gaussian probability density function. 

Substituting $a=0$ and $b=0$ into \eqref{derivative_1} and \eqref{derivative_2}, we have
\begin{align}
\left.\frac{\partial\mathcal{L}(\y;\bth_r)}{\partial \bth_r}\right|_{\bth_r=\bth_{r,0}} &=\sqrt{\frac{2}{\pi}}\[\begin{array}{c}
\sum_{i=1}^{N}\(r_iu_i + s_iv_i\)\\
\sum_{i=1}^{N}\(s_iu_i-r_iv_i\)
\end{array}\]\Nn\\
&=\sqrt{\frac{2}{\pi}}\[\begin{array}{c}
\text{Re}(\z^H\y)\\
\text{Im}(\z^H\y)
\end{array}\].
\label{derivative}
\end{align}
In addition, using the derivatives in \eqref{derivative_1} and \eqref{derivative_2}, we can compute the FIM in \eqref{FIM_definition}, element by element, as follows
\begin{align}
\F_{1,1}(\bth_{r,0})&=\frac{2}{\pi}\mathbb{E}\[\(\sum_{i=1}^{N}r_iu_i + \sum_{i=1}^{N}s_iv_i\)^2\]\Nn\\
&=\frac{2}{\pi}\mathbb{E}\[\sum_{i=1}^{N}s_i^2v_i^2+\sum_{i=1}^{N}r_i^2u_i^2\]
=\frac{2}{\pi}\tr(\Z\Z^H).
\end{align}
Similarly, we have
\begin{equation}
\F_{2,2}(\bth_{r,0})=\frac{2}{\pi}\tr(\Z\Z^H),
\end{equation}
and
\begin{align}
\F_{1,2}(\bth_{r,0})&=\frac{2}{\pi}\mathbb{E}\[\(\sum_{i=1}^{N}s_iv_i+\sum_{i=1}^{N}r_iu_i\)\right.\Nn\\
& \hspace{1.5cm} \left.\times\(\sum_{i=1}^{N}s_iu_i-\sum_{i=1}^{N}r_iv_i\)\]\Nn\\
&=\frac{2}{\pi}\mathbb{E}\[\sum_{i=1}^{N}s_i^2v_iu_i-\sum_{i=1}^{N}r_i^2v_iu_i\] =0,
\end{align}
which yields
\begin{equation}
  \label{FIM}
\F(\bth_{r,0})=\frac{2}{\pi}\tr(\Z\Z^H)\I_2,
\end{equation}
where $\I_2$ is the $2 \times 2$ identity matrix.

Substituting \eqref{derivative} and \eqref{FIM} into \eqref{Rao_definition}, the statistic becomes
\begin{align}
T_{\text{R}}=&\frac{|\tr(\Y\Z^H)|^2}{\tr(\Z\Z^H)},
\end{align}
and the proof follows by noting that
\begin{multline}
  \tr(\Z\Z^H) = \tr\left(\mathbf{a}_{r}(\phi) \mathbf{a}_{t}^{H}(\phi) \mathbf{S} \mathbf{S}^{H} \mathbf{a}_{t}(\phi) \mathbf{a}_{r}^{H}(\phi)\right) \\ = \| \mathbf{a}_{t}(\phi) \|^2 \| \mathbf{a}_{r}(\phi) \|^2\tr(\S\S^H) = N,
\end{multline}
where we have used $\| \mathbf{a}_{t}(\phi) \|^2 = p$, $\| \mathbf{a}_{r}(\phi) \|^2 = m$ and $\tr(\S\S^H)=n/p$.

\section{Proof of Theorem \ref{theorem:1} \label{appendix:B}}

For this proof, we need the following lemma, which is a multivariate version of the central limit theorem~\cite{Bentkus2005}.
\begin{lemma}\label{MCLT}
Let $\mathbf{s}=\sum_{i=1}^{N} \mathbf{q}_{i}$, where $\q_1,\cdots,\q_N \in \mathbb{R}^{d}$ are mutually independent random vectors with zero mean. 
Then, as $N \rightarrow \infty,$ $\mathbf{s}$ is asymptotically Gaussian distributed with zero mean and covariance matrix $\mathbf{C}$ if
\begin{align}\label{condition}
\lim _{N \rightarrow \infty} \sum_{i=1}^{N} \mathbb{E}\[\left\|\mathbf{C}^{-1 / 2} \mathbf{q}_{i}\right\|^{3}\]=0.
\end{align}
\end{lemma}

First, we shall compute the first and second order statistics and then study whether \eqref{condition} holds. For notational simplicity, let us define
\begin{align}
t_1=&\sqrt{N}w_1=\sum_{i=1}^{N} r_i u_i + \sum_{i=1}^{N} s_i v_i, \\
t_2=&\sqrt{N}w_2=\sum_{i=1}^{N} s_i u_i -\sum_{i=1}^{N} r_i v_i.
\end{align}
The first order moments of $t_1$ and $t_2$ are given by
\begin{align}
\mathbb{E}[t_1]=&\sum_{i=1}^{N} c_i u_i + \sum_{i=1}^{N} d_i v_i,\\
\mathbb{E}[t_2]=&\sum_{i=1}^{N} d_i u_i - \sum_{i=1}^{N} c_i v_i.
\end{align}
where
\begin{align}
c_i&=\mathbb{E}(r_i)=1-2Q(au_i-bv_i), \\
d_i&=\mathbb{E}(s_i)=1-2Q(av_i+bu_i).
\end{align}

For the second order moments, we can first compute the following expectations:
\begin{align}
\mathbb{E}[r_i r_j] &= \begin{cases}
c_i c_j, & i\neq j, \\
1, & i=j,
\end{cases}\\
\mathbb{E}[s_i s_j] &= \begin{cases}
d_i d_j, & i\neq j, \\
1, & i=j,
\end{cases}\\
\mathbb{E}[r_i s_j]&=c_id_j.
\end{align}
Then, we have
\begin{multline}
\mathbb{E}[t_1t_2] = -\sum_{i=1}^{N} c_i u_i \sum_{i=1}^{N} c_i v_i +\sum_{i=1}^{N} c_i u_i \sum_{i=1}^{N} d_i u_i \\
-\sum_{i=1}^{N} d_i v_i \sum_{i=1}^{N} c_i v_i + \sum_{i=1}^{N} d_i v_i \sum_{i=1}^{N} d_i u_i \\
-\sum_{i=1}^{N}(1-c_i^2)u_iv_i +\sum_{i=1}^{N}(1-d_i^2)u_iv_i \\
= \mathbb{E}[t_1]\mathbb{E}[t_2]+\sum_{i=1}^{N}(c_i^2-d_i^2)u_iv_i,
\end{multline}
and
\begin{align}
\mathbb{E}[t_1^2]
&=\mathbb{E}^2[t_1]+\tr(\Z\Z^H)-\sum_{i=1}^{N} \left(c_i^2 u_i^2 + d_i^2 v_i^2 \right),\\
\mathbb{E}[t_2^2]&=\mathbb{E}^2[t_2]+\tr(\Z\Z^H)-\sum_{i=1}^{N}\left(c_i^2 v_i^2 + u_i^2d_i^2\right).
\end{align}
Combining all results above and using the fact that $\tr(\Z\Z^H)=N$, we obtain the first and second order statistics of $\w$ in \eqref{w_mean} and \eqref{w_covariance}.

To conclude the derivation, we prove that \eqref{condition} holds. Define $\q_i$ as
\begin{equation}
\q_i = \frac{1}{\sqrt{N}}\begin{bmatrix}
u_ir'_i+v_is'_i\\
u_is'_i-v_ir'_i
\end{bmatrix}, \quad i = 1, \ldots, N,
\end{equation}
where
\begin{align}
r'_i &= r_i-c_i, &
s'_i &= s_i-d_i.
\end{align}
Then, $\mathbf{s}=\sum_{i=1}^{N} \mathbf{q}_{i}$ in Lemma \ref{MCLT} corresponds to $\w$ in \eqref{w} and $\C=\bSi_w (\beta)$. It is shown in~\cite{Levin2011TIT} that
\begin{align}
\sum_{i=1}^{N} \mathbb{E}\left\|\mathbf{C}^{-1 / 2} \mathbf{q}_{i}\right\|^{3} \leq\left\|\mathbf{C}^{-1 / 2}\right\|_\text{F}^{3}  \ \sum_{i=1}^{N}\left(\mathbb{E}\left\|\mathbf{q}_{i}\right\|^{4}\right)^{3 / 4},
\end{align}
and since 
$\left\|\mathbf{C}^{-1 / 2}\right\|_\text{F}^{3}$ is bounded, a sufficient condition for \eqref{condition} is
\begin{align}\label{condition_2}
\lim_{n\rightarrow\infty} \sum_{i=1}^{N}\left(\mathbb{E}\left\|\mathbf{q}_{i}\right\|^{4}\right)^{3 / 4}=0.
\end{align}
To proceed, we further expand $\left\|\mathbf{q}_{i}\right\|^{4}$ as
\begin{align}
\left\|\mathbf{q}_{i}\right\|^{4}=\[\frac{(u_i^2+v_i^2)(r_i'^2+s_i'^2)}{N}\]^2.
\end{align}
Recalling that $r_i,s_i\in\{\pm 1\}$, we have the following upper bounds:
\begin{align}
\mathbb{E}\[r_i'^4\] &\leq \frac{4}{3}, &
\mathbb{E}\[s_i'^4\] &\leq \frac{4}{3}, &
\mathbb{E}\[r_i'^2s_i'^2\] &\leq 1,
\end{align}
which yield
\begin{align}\label{condition_3}
\sum_{i=1}^{N}\left(\mathbb{E}\left\|\mathbf{q}_{i}\right\|^{4}\right)^{3 / 4}\leq \(\frac{14}{3}\)^{\frac{3}{4}} \sum_{i=1}^{N} \frac{(u_i^2+v_i^2)^{3/2}}{N^{3/2}}.
\end{align}
Since the right-hand-side 
of \eqref{condition_3} is a sum of $N$ terms that are of order $N^{-\frac{3}{2}}$, it approaches $0$ as $N\rightarrow \infty$, which completes the proof.

%

\section{Derivation of \eqref{non_centrality_parameter}}
\label{appendix:D}

In the $\infty-$bit case, the parameter space is $\bth=[a,b,\sigma_n^2]^T$, where $\sigma_n^2$ is a nuisance parameter and the relevant parameters are collected in $\bth_r=[a,b]^T$.

By dropping the constant terms in the log-likelihood function, it becomes
\begin{align}
\mathcal{L}(\X;\bth_r,\sigma_n^2)=-N\ln(\sigma_n^2)-\frac{||\X-\beta\Z||^2_\text{F}}{\sigma_n^2}.
\end{align}
The FIM can be computed as:
\begin{align}
\F&=-\mathbb{E}\[\left.\frac{\partial^2 \mathcal{L}}{\partial \bth^2} \right| \bth\]\Nn\\
&=\mathbb{E}
\begin{bmatrix}
\frac{2N}{\sigma_n^2} & 0 & \frac{2h-2Na}{\sigma_n^4}  \\
0 & \frac{2N}{\sigma_n^2} & -\frac{2g+2Nb}{\sigma_n^4} \\
\frac{2h-2Na}{\sigma_n^4} & -\frac{2g+2Nb}{\sigma_n^4} & \frac{2||\X-\beta\Z||^2_F}{\sigma_n^6}-\frac{N}{\sigma_n^4}
\end{bmatrix},
\end{align}
where
\begin{align}
h&=\operatorname{Re}[\tr(\X^H\Z)], &
g&=\operatorname{Im}[\tr(\X^H\Z)].
\end{align}
Now, taking into account that
\begin{align}
\mathbb{E}[||\X-\beta\Z||^2_\text{F}]&=N\sigma_n^2, &
\mathbb{E}[h]&=Na, &
\mathbb{E}[g]&=-Nb,
\end{align}
the FIM becomes
\begin{equation}
\F =
\begin{bmatrix}
\frac{N}{\sigma_n^4} & 0 & 0 \\
0 & \frac{2N}{\sigma_n^2} & 0 \\
0 & 0 & \frac{2N}{\sigma_n^2}
\end{bmatrix}.
\end{equation}
Therefore, we have
\begin{equation}
\[\F^{-1}(\bth_{1})\]_{\bth_{r},\bth_{r}}^{-1}=\frac{2N}{\sigma_n^2}\I_2,
\end{equation}
and the non-centrality parameter in \eqref{non_centrality_parameter} can be computed as
\begin{align}
\delta^2 &=(\bth_{r_1}-\bth_{r_0})^T\[\F^{-1}(\bth_{1})\]_{\bth_{r},\bth_{r}}^{-1}(\bth_{r_1}-\bth_{r_0}) \Nn \\
&=N\frac{2|\beta|^2}{\sigma_n^2},
\end{align}
where $\[\F^{-1}(\bth_{1})\]_{\bth_{r},\bth_{r}}$ is the block of the inverse FIM corresponding to the parameters in $\bth_r$.
Recalling that we have set $\sigma_n^2=2$ in Section \ref{sec:detector}, we obtain $\delta^2=N|\beta|^2,$ which concludes the derivation of \eqref{non_centrality_parameter}.

\section{Proof of Theorem \ref{th:nonnull_approx_Rao}}
\label{appendix:E}

Since we have assumed that $\beta$ is of order $\mathcal{O}(N^{-\frac{1}{2}})$, where $N=mn$, we can apply a Taylor's approximation to the $Q$ function around $0$, allowing us to write
\begin{align}
c_i=1-2Q(au_i-bv_i) = \sqrt{\frac{2}{\pi}}(au_i-bv_i)+\mathcal{O}(N^{-1}), \\
d_i=1-2Q(av_i+bu_i) = \sqrt{\frac{2}{\pi}}(av_i+bu_i)+\mathcal{O}(N^{-1}).
\end{align}
Then, $\sigma_i^2$ and $\sigma_{12}$ become
\begin{multline}
\sigma_1^2 = 1-\frac{2}{N\pi}\sum_{i=1}^{N}\[a^2(u_i^4+v_i^4)+2ab(u_iv_i^3-u_i^3v_i)\right. \\
\left.+2b^2u_i^2v_i^2\]+\mathcal{O}(N^{-2}),
\end{multline}
\begin{multline}
\sigma_2^2 = 1-\frac{2}{N\pi}\sum_{i=1}^{N}\[b^2(u_i^4+v_i^4)+2ab(u_i^3v_i-u_iv_i^3)\right. \\
\left.+2a^2u_i^2v_i^2\]+\mathcal{O}(N^{-2}),
\end{multline}
and
\begin{multline}
\sigma_{12} = \frac{2}{N\pi}\sum_{i=1}^{N}\[(a^2-b^2)(u_i^2-v_i^2)-4abu_iv_i\]u_iv_i \\
+\mathcal{O}(N^{-2}).
\end{multline}
Recalling again that $a$ and $b$ are of order $\mathcal{O}(N^{-\frac{1}{2}})$, the covariance matrix of $\w=[w_1,w_2]^T$ becomes
\begin{equation}
\bSi_w = \I_2 + \mathcal{O}(N^{-1}).
\end{equation}
Then, the weighted sum in \eqref{eq:Rao_weightedsum} has the identical weigths as the eigenvalues of $\bSi_w(\beta)$ becomes $\lambda_i = 1 + \mathcal{O}(N^{-1})$. Additionally, the means of $w_1$ and $w_2$ can be approximated as:
\begin{align}
\mathbb{E}[w_1] &=  \sqrt{\frac{2}{N\pi}}a\sum_{i=1}^{N}(u_i^2+v_i^2)+\mathcal{O}(N^{-\frac{1}{2}}) \Nn\\
&= \sqrt{\frac{2N}{\pi}}a+\mathcal{O}(N^{-\frac{1}{2}}),
\end{align}
and
\begin{align}
\mathbb{E}[w_2] &=  \sqrt{\frac{2}{N\pi}}b\sum_{i=1}^{N}(u_i^2+v_i^2)+\mathcal{O}(N^{-\frac{1}{2}}) \Nn\\
&= \sqrt{\frac{2N}{\pi}}b+\mathcal{O}(N^{-\frac{1}{2}}).
\end{align}
Therefore, $T_{\text{R}}=w_1^2+w_2^2$ is the sum of squares of two uncorrelated Gaussian random variables with means $a\sqrt{2N/\pi}$ and $b\sqrt{2N/\pi}$ and unit variance, which results in $T_{\text{R}}$ that follows a non-central $\chi^2$ distribution with DOF 2 and non-centrality parameter
\begin{align}
\delta_1^2 = \frac{2N}{\pi}|\beta|^2.
\end{align}
This completes the proof of Theorem \ref{th:nonnull_approx_Rao}.

\bibliographystyle{IEEEtran}


\end{sloppypar}


\end{document}